\documentclass[11pt,twoside]{article}
\usepackage{cal05}

\contact{Soeren Larsen}
\email{slarsen@eso.org}

\begin{document}

%
%
%

\title{Calibration of ACS Prism Slitless Spectroscopy Modes}
\titlemark{ACS Prism Calibrations}

%

\author{S. S. Larsen, M. K{\"u}mmel and J. R. Walsh}
\affil{ESO/ST-ECF, Karl-Schwarzschild-Str. 2, D-85748 Garching bei
    M{\"u}nchen, Germany}

%
%

\paindex{Larsen, S. S.}
\aindex{K{\"u}mmel, M.}
\aindex{Walsh, J. R.}

%
%

\authormark{Larsen, K{\"u}mmel \& Walsh}



\begin{abstract}
The Advanced Camera for Surveys is equipped with three prisms in the Solar 
Blind (SBC) and High Resolution (HRC) Channels, which together cover the 
1150 -- 3500 \AA\ range, albeit at highly non-uniform spectral resolution.
We present new wavelength- and flux calibrations of the SBC (PR110L and 
PR130L) and HRC (PR200L) prisms, based on calibration observations obtained 
in Cycle 13.  The calibration products are available to users via the
ST-ECF/aXe web pages, and can be used directly with the aXe package.  We 
discuss our calibration strategy and some caveats specific to slitless prism 
spectroscopy.
\end{abstract}




\section{Introduction}

The Advanced Camera for Surveys is currently the only instrument on HST
which provides spectroscopy in the UV-optical range.  With STIS being
unavailable (at least for the moment), the interest in the ACS spectroscopic 
modes has increased substantially. An extensive 
calibration effort was undertaken in Cycle 13 in order to provide much 
improved wavelength- and flux calibrations for the prism modes which had 
seen little use in earlier cycles. The ACS has two prisms (PR110L and PR130L) 
installed in the Solar Blind Channel (SBC), both covering the wavelength range 
from roughly 1200\AA\ -- 2000\AA . The main difference between the two
SBC prisms is that the sensitivity of PR110L extends below the geocoronal
Ly $\alpha$, while that of the PR130L does not. The sensitivity below
1216 \AA\ results in a significantly
higher sky background for the PR110L.  The one prism (PR200L) in the 
High Resolution Channel covers the range $\sim1800$\AA -- 3500\AA . In 
this paper we present the calibration observations and discuss the 
derived trace-, wavelength and flux calibrations. The calibrations 
presented here are implemented in configuration files for the aXe package 
(K{\"u}mmel et al., these proceedings) and are made available to users via 
the aXe web pages ({\tt http://www.stecf.org/software/axe}). 


Spectroscopic observations with the ACS prisms share many similarities with 
the G800L grism in the WFC and HRC channels, but there are also some 
differences. In 
both cases the object spectrum (or spectra) may fall anywhere on the detector, 
and the reference point for the wavelength scale is typically established 
using a direct image taken 
immediately before (or after) the prism exposure. The direct image is even
more crucial for the prism modes, which do not show a zeroth order
spectrum (nor any higher- or negative orders).  Contrary to the case of 
grism spectroscopy, the wavelength scale of the prism spectra is highly 
non-linear, with the spectral resolution decreasing towards longer 
wavelengths (see below). For the PR200L, this causes a ``red pile-up'' with 
wavelengths between 4000 \AA\ and 10000 \AA\ being compressed into only 7 
pixels.  This pile-up can be particularly troublesome for red objects,
where light from the diffraction spikes and outer halo of the PSF may
contaminate the bluer parts of the spectrum. These effects are still 
poorly quantified.  No such pile-up is seen for the SBC prisms, due to 
the lack of sensitivity at wavelengths $>2000$\AA\ for the MAMA detector. 

\section{Observations}

  The relatively low spectral resolution of the prisms, particularly at
longer wavelengths, and the high spatial resolution of the SBC and HRC
cameras limit the choice of suitable wavelength calibrators. The calibration
targets must have strong emission features in the ultraviolet spectral
region, and at the same time be compact.  For the SBC, an additional
constraint was the ``bright object protection'' limit.  We eventually settled 
on a combination of a planetary nebula in the Large Magellanic Cloud 
(LMC-SMP-79) and a QSO at redshift $z=0.836$ (Q0037-3544) for the PR200L, 
and two QSOs at redshifts $z=0.168$ (PG1322$+$659) and $z=0.098$ 
(PG1404$+$226) for the SBC prisms.  The PN was known from existing STIS 
spectra to have strong C III] and [C II] emission lines and be relatively 
compact (though not point-like).  The QSOs were selected from the 
compilation of V{\'e}ron-Cetty \& V{\'e}ron (2003), which lists 48921 quasars
and makes it possible to select a QSO with emission features
(Ly $\alpha$, C IV etc.) at essentially any desired wavelength. We required the
calibration targets to have existing HST spectra (either from FOS or STIS)
and low reddenings ($A_B < 0.1$ mag), but even with these additional 
constraints it was not difficult to find suitable QSO targets.
For the flux calibration we observed two white dwarf standards, 
WD1657$+$343 and LDS749B. STIS spectra of these stars were kindly
provided by R.\ Bohlin.

Each target was observed at several positions across the detector in order
to map spatial variations in the trace- and wavelength solutions, as well
as any large-scale sensitivity variations. For each prism, one wavelength 
calibration target was observed at 9 or 10 positions (i.e.\ 2 orbits) and the 
other at 5 positions (one orbit). The flux standards were observed at 5--6 
different positions. Each prism exposure was preceded by a direct image 
through the F165LP filter (for the SBC) or F330W (for the HRC). The F165LP 
filter was chosen in order to avoid conflict with the bright object protection
limit for the QSO targets. However, for one of the flux standards we
obtained direct images in both the F122M and F165LP filters in order to 
check for any offsets between exposures in the two bands.  Such offsets
were indeed found (see below).

\begin{figure}
\epsscale{0.40}
\centerline{\framebox{\plotone{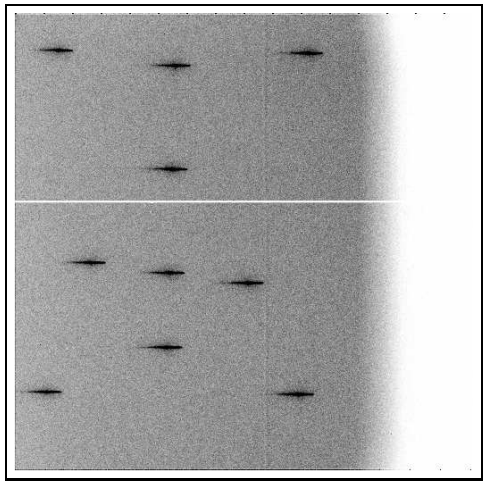}}}
\caption{Sum of the 10 pointings obtained for PG1322$+$659 for the PR110L. 
Note the vignetting at the right-hand side of the field. A few dead
rows at $Y\approx600$ are seen as a horizontal white line.}
\label{fig:fovsbc}
\end{figure}

The coverage of the field is illustrated in Fig.~\ref{fig:fovsbc}, which
shows a sum of the 10 PR110L exposures obtained for PG1322$+$659. The
MAMA detector has a few dead rows at $Y\approx600$ which show up as a
horizontal white line in the figure. To avoid these, the default aperture
center is at (512, 400).
Note that there is significant vignetting at the
right-hand side of the field, affecting $\sim$200 image columns. A similar
phenomenon is seen for the HRC/PR200L, but affecting the left-hand side
of the field. 

\begin{figure}
\plottwo{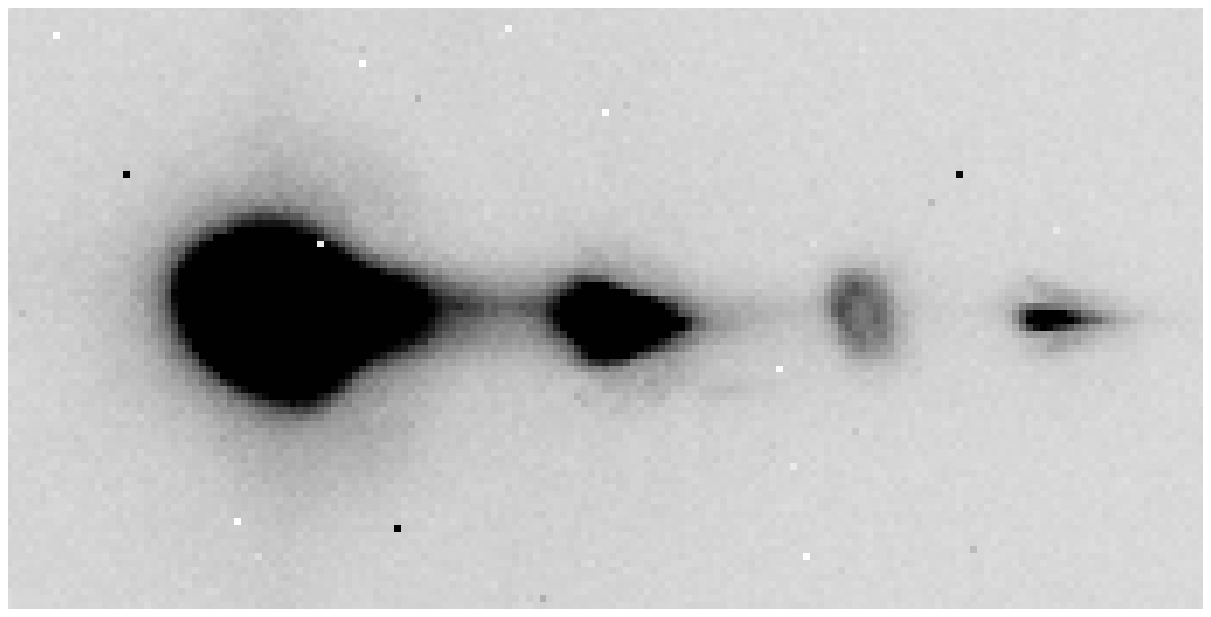}{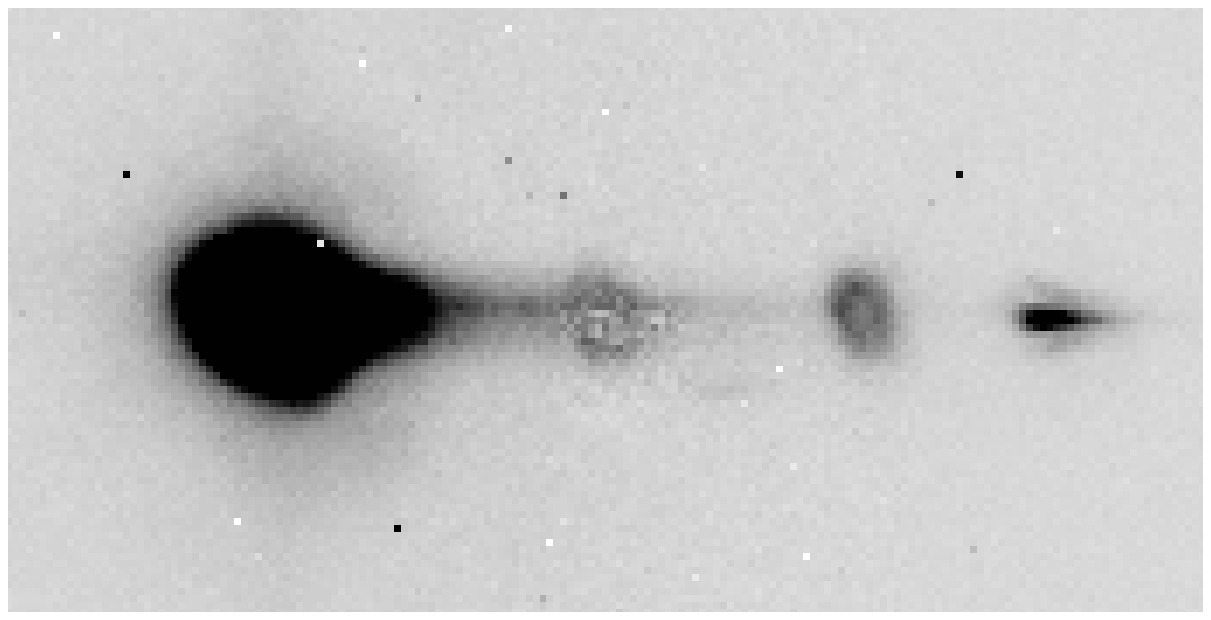}
\caption{PR200L image of LMC-SMP-79. Left: raw image, Right: after subtraction
 of contaminating object. Note that the nebula is resolved at the HRC
 resolution. From the left to the right, the four main ``features'' in the
 spectrum are: the red pile-up, [C II] 2325\AA / contaminating star,
 C III] 1909\AA, and another contaminating star.}
\label{fig:pr200l_contam}
\end{figure}

A few difficulties were encountered with the data for LMC-SMP-79. First,
with a diameter of about $0\farcs3$, the planetary nebula was slightly 
resolved at the resolution of the HRC. Second, by an unfortunate
coincidence the prism spectra of
the nebula were contaminated by two stars in the field, one of which 
affected the [C II] 2325 \AA\ line (Fig.~\ref{fig:pr200l_contam}).
Fortunately, we found that the contaminating object could be removed 
by scaling and subtracting the spectrum of a star of similar spectral
type located elsewhere in the field. This procedure worked well enough
that we could recover the [C II] 2325 \AA\ line. Fig.~\ref{fig:pr200l_contam}
also illustrates the PR200L ``red pile-up'', which is the dominant feature
at the left (red) end of the spectrum.


\section{Wavelength- and trace solutions}

The spectral trace- and wavelength solutions are defined with respect to
a reference position, ($X_{\rm ref}, Y_{\rm ref}$), which was
measured by running the SExtractor code (Bertin \&
Arnouts 1996) on the direct images.  The trace 
descriptions were then derived by measuring the centroid along the image
columns (corresponding to the spatial direction) for the flux standard 
spectra, and fitting a straight line
(for the PR110L and PR200L) or 2nd order polynomial (for PR130L) to
the offset $\Delta Y = Y_{\rm trace} - Y_{\rm ref}$ vs.\
$\Delta X = X_{\rm trace} - X_{\rm ref}$.  
In all cases the dispersion direction of the prism spectra was found to
be aligned with the image rows to within
1 degree, towards positive $\Delta X$ for PR200L and negative $\Delta X$
for the SBC prisms.  The variation in the offsets and slopes 
$\Delta Y / \Delta X$ of the prism spectra was fitted with 2-dimensional 
1st order polynomials as a function of position on the detector (not
reproduced here due to space constraints).

\begin{table}
\caption{Wavelength solution terms}
\label{tab:wsol}
\scriptsize
\begin{center}
\begin{tabular}{lcc} \tableline \\ 
SBC:   &  PR110L & PR130L \\
 $a_0$    &  $-95.70 - 0.00662 \times X_{\rm ref} + 0.01760 \times Y_{\rm ref}$ &
          $-80.61 - 0.00711 \times X_{\rm ref} + 0.01547 \times Y_{\rm ref}$ \\
 $a_1$    & $1049.39 + 0.02105 \times X_{\rm ref} - 0.01867 \times Y_{\rm ref}$ &
         $1072.71 + 0.01249 \times X_{\rm ref} - 0.01699 \times Y_{\rm ref}$ \\
 $a_2$    & 2.11506$\times10^4$    & -1.49676$\times10^4$ \\
 $a_3$    & 9.23749$\times10^6$    &  1.94800$\times10^6$ \\
 $a_4$    & 4.93403$\times10^8$    &  5.48481$\times10^7$ \\
 $a_5$    & 1.24585$\times10^{10}$ &  1.11636$\times10^9$ \\ \tableline \\
HRC:   &  PR200L \\
 $a_0$    &  $145.755 + 0.00872 \times X_{\rm ref} - 0.01925 \times Y_{\rm ref}$ \\
 $a_1$    & $1147.24  - 0.03042 \times X_{\rm ref} - 0.00782 \times Y_{\rm ref}$ \\
 $a_2$    & 8.11002$\times10^4$ \\
 $a_3$    & -8.28929$\times10^5$ \\
 $a_4$    & -4.65350$\times10^6$ \\
 $a_5$    &  1.68888$\times10^8$ \\ \tableline
\end{tabular}
\end{center}
\end{table}

Wavelength solutions were assumed to be of the form used by
Bohlin et al.\ (2000), i.e.\
\begin{equation}
  \lambda =   a_1 
            + \frac{a_2}{(\Delta X-a_0)}
	    + \frac{a_3}{(\Delta X-a_0)^2}
	    + \frac{a_4}{(\Delta X-a_0)^3}
	    + \frac{a_5}{(\Delta X-a_0)^4}
\end{equation}
where each of the coefficients $a_0 \ldots a_5$ can be a function of 
($X_{\rm ref}, Y_{\rm ref}$). In order to define the wavelength solutions,
the prism spectra were first extracted with aXe using a configuration
file where the ``wavelength'' scale was simply the pixel offset,
$\Delta X$, along
the trace. The aXe spectra were then converted to IRAF format, and the
$\Delta X$ values for the various emission lines were measured with
the SPLOT task, using a Gaussian fit. We then solved directly for the
full, spatially dependent, wavelength solution. 
In the PN spectrum we could measure the C III] 1909\AA\ and [C II] 2325\AA\
lines and in the spectrum of Q0037-3544 we could measure Ly$\alpha$,
C IV and C III] redshifted to 2233\AA, 2844\AA\ and 3503\AA, thus
providing 5 wavelength sampling points for the PR200L. 
For the SBC prisms, we detected Ly $\alpha$ redshifted to 1420\AA\ and
1335\AA\ in the spectra of PG1322$+$659 and PG1404$+$226 respectively,
as well as C IV redshifted to 1701\AA\ for PG1404$+$226. The spectra of
PG1322$+$659 had too low resolution at the wavelength of C IV (1806\AA)
to allow useful measurements of this feature. Thus, we have 3 wavelength
sampling points for the calibration of the SBC prisms.
The resulting wavelength solutions are reproduced in Table~\ref{tab:wsol}.
Note that we have adopted the higher order terms from Bohlin et al.\ (2000),
i.e. $a_3$, $a_4$ and $a_5$ for the PR200L, and $a_2\ldots a_5$ for
PR110L and PR130L. However, the $a_2$ term we derive for the PR200L 
is quite similar to that derived by Bohlin et al. (81100 vs.\ 83000).

\begin{figure}
\plottwo{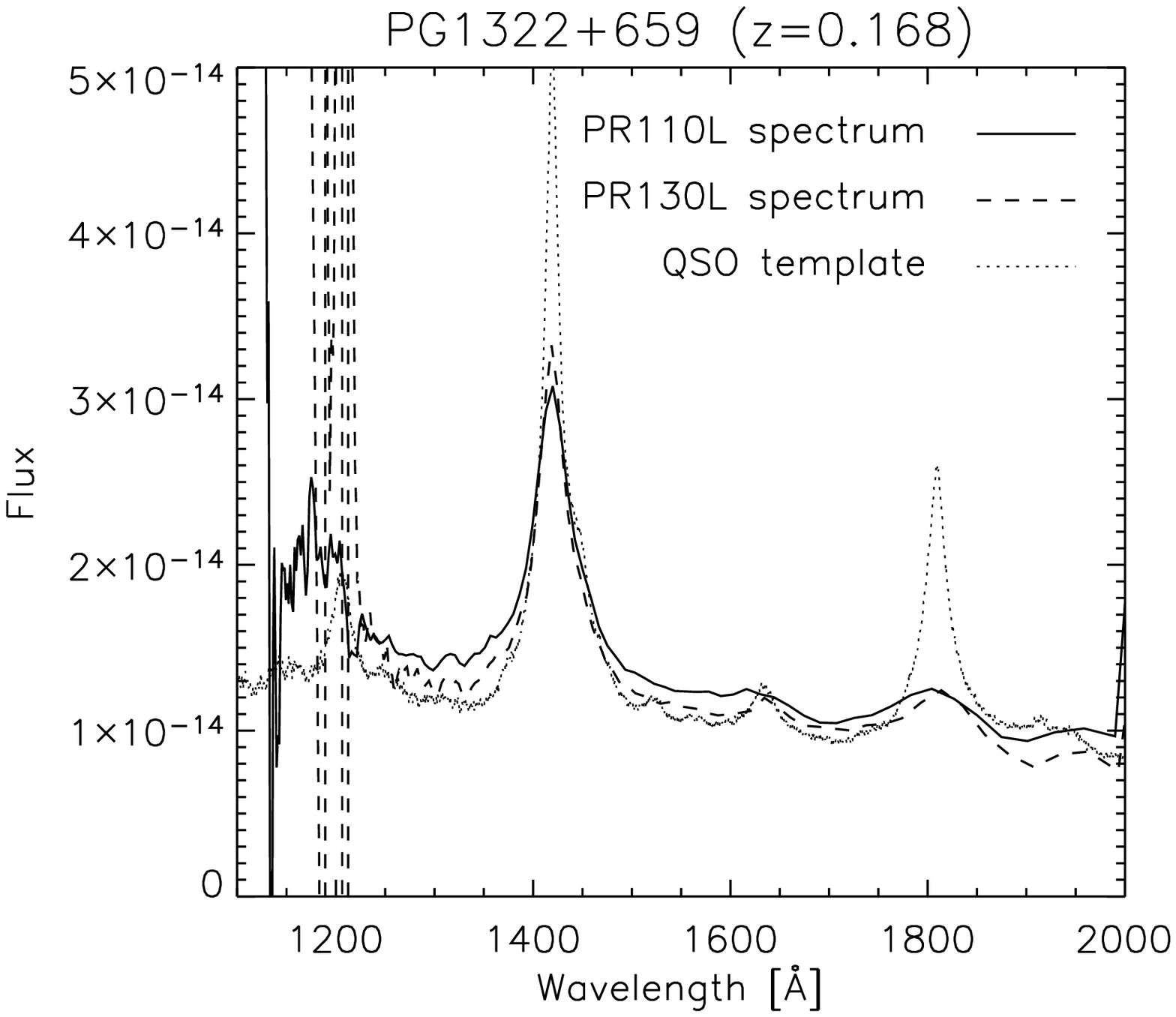}{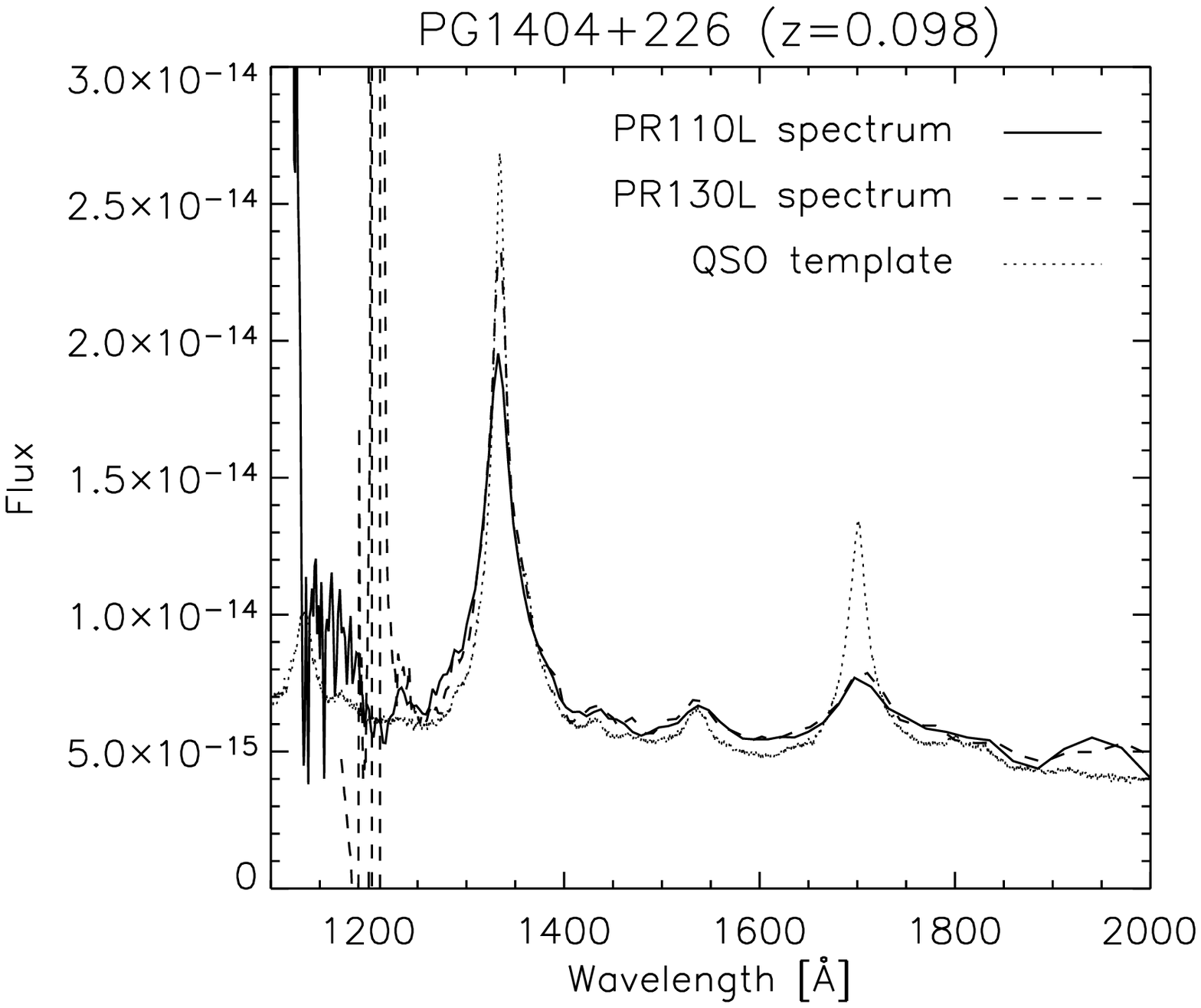}
\caption{SBC PR110L and PR130L spectra of PG1322$+$659 (left) and
 PG1404$+$226 (right). A scaled template QSO spectrum (from Zheng et 
 al.\ 1997) is shown for comparison.}
\label{fig:qsos_sbc}
\end{figure}

\begin{figure}
\plottwo{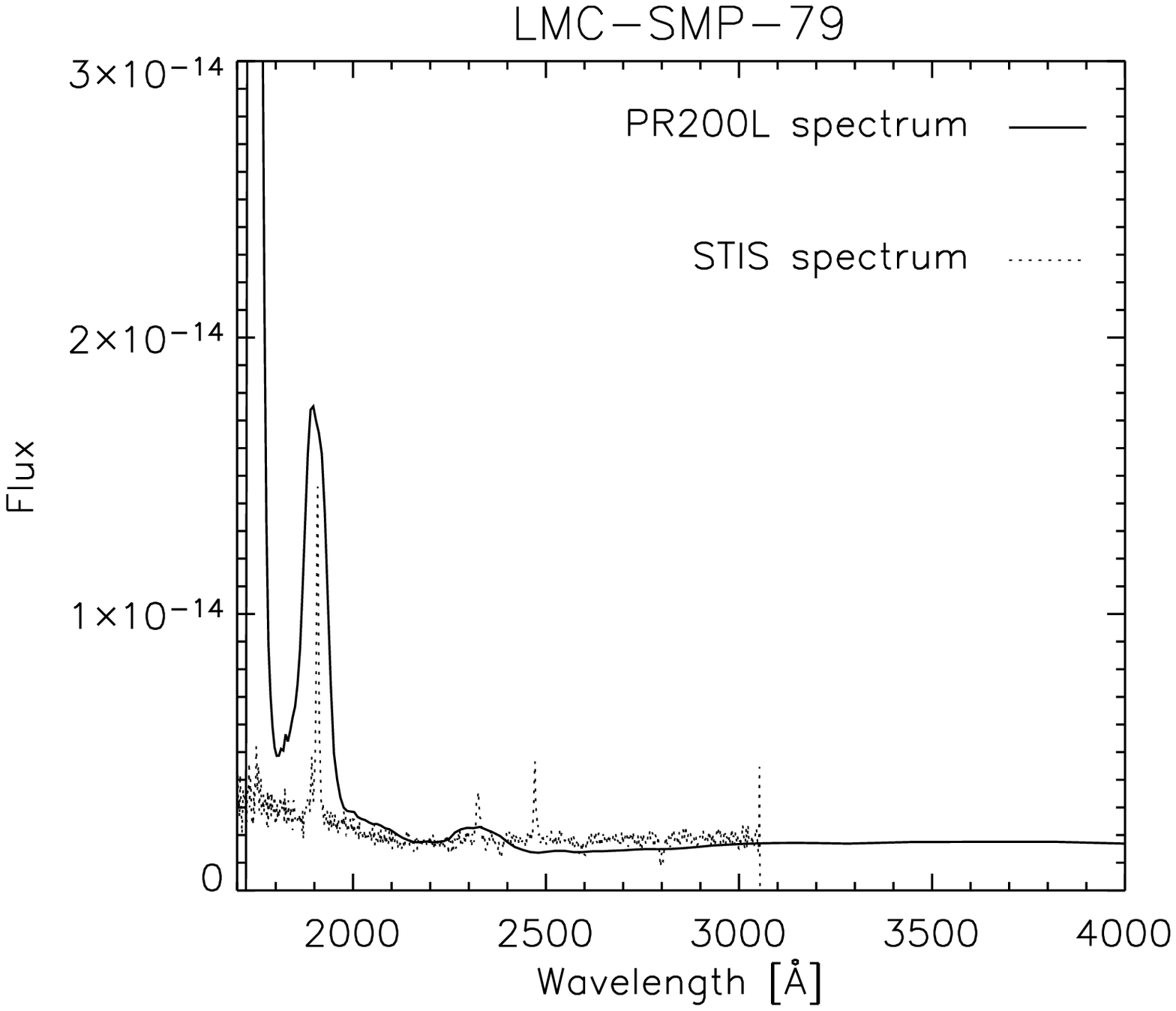}{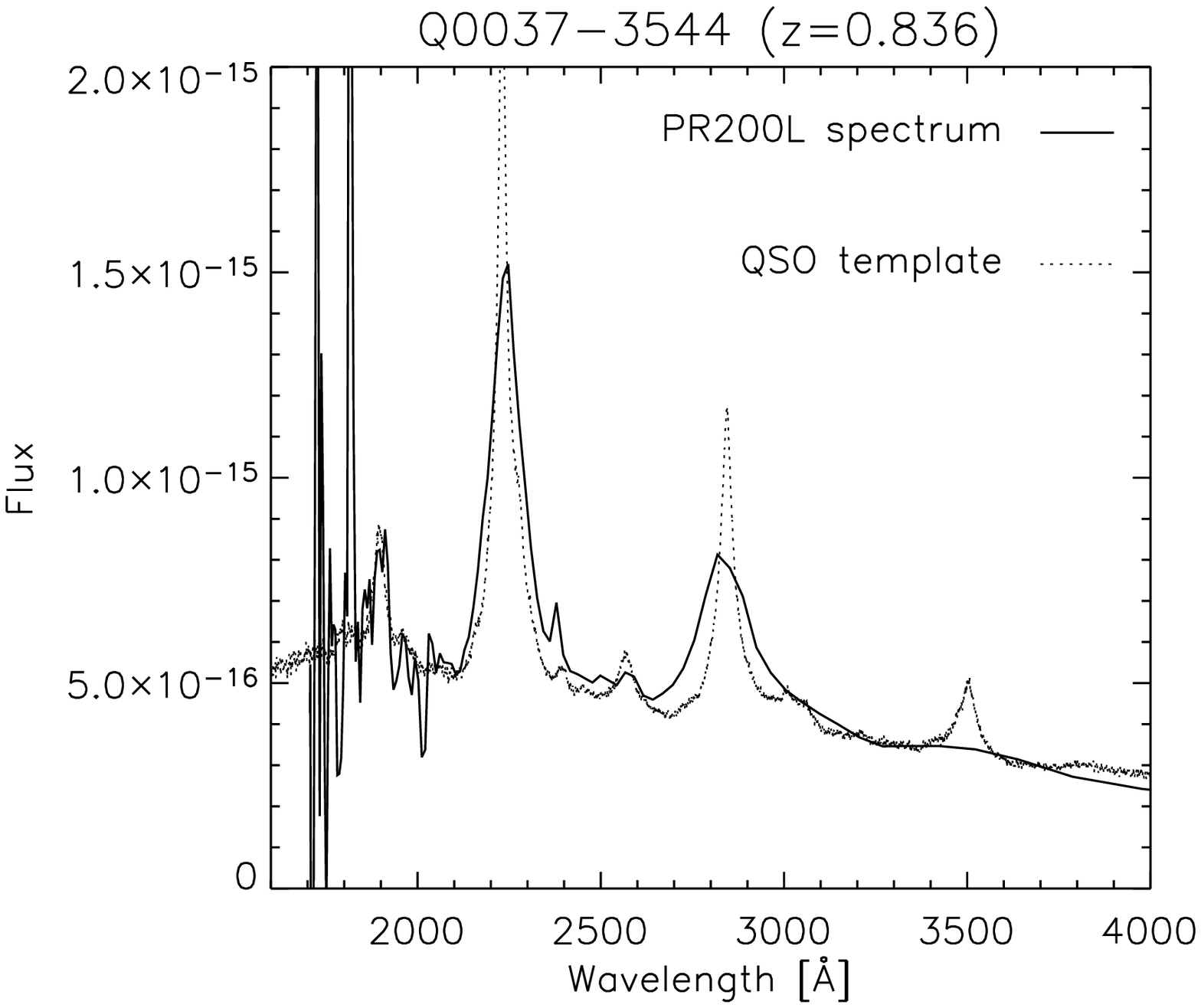}
\caption{HRC PR200L spectra of LMC-SMP-79 (left) and the quasar
 Q0037-3544 (right). For comparison, a STIS spectrum of LMC-SMP-79 
 (courtesy L. Stanghellini) and a scaled quasar template spectrum are shown.
 }
\label{fig:pn_qso_hrc}
\end{figure}

The trace- and wavelength solutions derived here assume that direct
images are taken in the F165LP filter (for the SBC) or in F330W (for the
HRC). For the SBC, images taken in different filters show offsets of
up to 12 pixels, so it is important to correct any coordinates measured
on images taken in filters other than F165LP for these offsets.
A preliminary analysis of these offsets, using our data and data from
program 9563 (P.I.: G.\ De Marchi), suggests that the following
corrections should be applied (all in pixels): 
F115LP: $(\Delta x, \Delta y) = (-3.74, 12.13)$; 
F122M: $(\Delta x, \Delta y) = (-3.74, 12.07)$; 
F125LP: $(\Delta x, \Delta y) = (0.97, -2.44)$;
F140LP: $(\Delta x, \Delta y) = (2.88, -8.48)$ and
F150LP: $(\Delta x, \Delta y) = (-0.31, 1.26)$.

Sensitivity functions were derived for each prism by dividing
the prism spectra of the flux standards with the standard STIS spectra, and
fitting the ratio with a spline function. The spectra of the flux standards
were extracted using an extraction box height of $\pm0\farcs5$. 
A flatfield cube (K{\"u}mmel et al., these proc.) has been defined for the 
SBC prisms, although not including any wavelength-dependency. This shows 
sensitivity variations across the SBC detector amounting to a few percent.

\begin{figure}
\epsscale{0.93}
\plotone{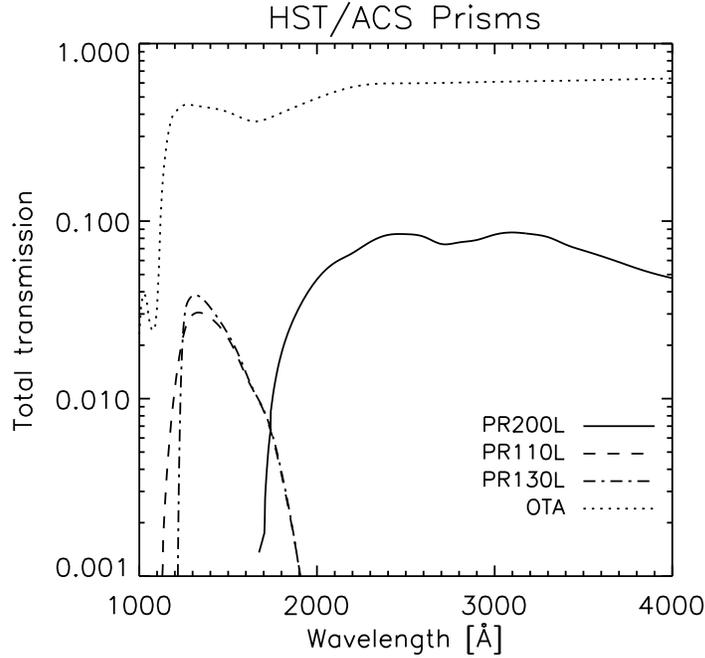}
\caption{Total transmission of the ACS prisms relative to the ``perfect
 telescope'', based on the sensitivity curves derived here. The dotted
 line indicates the throughput of the OTA.
 }
\label{fig:trans}
\end{figure}


\section{Consistency checks}

Figs.~\ref{fig:qsos_sbc} and \ref{fig:pn_qso_hrc} show the SBC
and HRC prism spectra of the wavelength calibration targets, extracted 
with aXe, compared
with a template QSO spectrum (Zheng et al.\ 1997) and the STIS spectrum
of LMC-SMP-79 (courtesy of L.\ Stanghellini). The QSO template spectrum
has been adjusted to the redshift of the corresponding targets and
scaled to match the data. No reddening correction has been applied, but
the observed QSOs all have $A_B < 0.1$ mag. The prism observations appear
in good agreement with the reference spectra, both as far as the
wavelength scales and flux calibrations are concerned.  While comparison of
aXe-extracted prism spectra of the \emph{flux} standards with the STIS 
reference spectra shows
agreement to about 5\%, this agreement may be somewhat deceptive, since the 
sensitivity files were of course derived from those observations in the
first place.  The comparison in Figs.~\ref{fig:qsos_sbc} and 
\ref{fig:pn_qso_hrc} thus provides a welcome independent check, for objects
which have rather different spectral energy distributions than the 
WD standards.  While the absolute scalings of the reference spectra here
are arbitrary, the overall shape of the prism spectra appears correct, even
at longer wavelengths where the wavelength solutions become increasingly
uncertain.  However, note that the decrease in spectral resolution at 
longer wavelengths makes it difficult to detect even relatively strong 
features, such as those seen in the QSO spectra. 

The wavelength scale is generally accurate to better than a pixel.  The 
dispersion of the SBC prisms is about 2 \AA\ pixel$^{-1}$ at 1200 \AA, but 
degrades to about 10 \AA\ pixel$^{-1}$ at 1600 \AA\ and 30 \AA\ pixel$^{-1}$ 
at 2000 \AA . The PR200L has a dispersion of about 5 \AA\ pixel$^{-1}$ 
at 1800 \AA\ and nearly 40 \AA\ pixel$^{-1}$ at 3000 \AA . So while the 
three prisms together cover the full wavelength range from 1150 \AA -- 
3500 \AA, the spectral resolution is highly non-uniform.

Using the sensitivity curves derived here, we can calculate the total
system throughput for each prism, as shown in Fig.~\ref{fig:trans}. The
PR200L reaches a peak efficiency of about 10\%, while the SBC prisms
peak at 3--4\%. These estimates are simply the measured count rates vs.\
the maximum possible rates, assuming a telescope aperture of 4.52 m$^{2}$,
and the throughput curves thus include all the effects of mirror coatings,
detector sensitivity etc. For comparison, the throughput of the Optical
Telescope Assembly (OTA) alone is also shown.

\section{Summary}

The SBC and HRC prisms are now fully calibrated, both in wavelength and
flux. The calibration products are available for use with the aXe
package, and can be downloaded from the aXe web pages.  Future calibration
observations will aim to monitor the stability of the prism modes and 
possibly provide various improvements, such as quantifying the effect of
scattered light from the red pile-up in the PR200L. A better understanding
of the general properties of the SBC and HRC detectors, such as CTE
effects, may also lead to improvements in prism spectroscopy.

\end{document}